\documentclass[osajnl,twocolumn,showpacs,superscriptaddress,10pt]{revtex4-1}

\usepackage{amsmath,amssymb,graphicx}
\usepackage{xcolor}
\usepackage{tikz}
\usepackage{adjustbox}
\usepackage{todonotes}

\usetikzlibrary{shapes.geometric, arrows}
\tikzstyle{input} = [trapezium, trapezium left angle=70, trapezium right angle=110, minimum width=3cm, minimum height=1cm, text centered, draw=black, fill=blue!30, text width=3.5cm]
\tikzstyle{process} = [rectangle, minimum width=3cm, minimum height=1cm, text centered, draw=black, fill=orange!30, text width=3.5cm]
\tikzstyle{decision} = [diamond, minimum width=3cm, minimum height=0.2cm, text centered, draw=black, fill=green!30, text width=2cm]
\tikzstyle{arrow} = [thick,->,>=stealth]

\begin{document}

\title{A new radiobiology-based HDR brachytherapy treatment planning algorithm used to investigate the potential for hypofractionation in cervical cancer}
\author{Kaelyn Seeley}
\affiliation{Department of Physics and Astronomy, University of Pittsburgh, Pittsburgh, PA}
\author{I-Chow J. Hsu}
\affiliation{Department of Radiation Oncology, University of California San Francisco, Comprehensive Cancer Center, San Francisco, CA}
\author{Tae Min Hong}
\affiliation{Department of Physics and Astronomy, University of Pittsburgh, Pittsburgh, PA}
\author{J. Adam Cunha}
\affiliation{Department of Radiation Oncology, University of California San Francisco, Comprehensive Cancer Center, San Francisco, CA}

\begin{abstract}
\setlength{\parindent}{0cm}
\underline{Purpose:} Most commercially available treatment planning systems for brachytherapy operate based on physical dose and do not incorporate fractionation or tissue-specific response. The purpose of the study is to investigate the potential for hypofractionation in HDR brachytherapy, thereby reducing the number of implants required.

\underline{Methods and Materials:} A new treatment planning algorithm was built in order to optimize based on tissue and fractionation specific parameters. Different fractionation schemes were considered for 6 patients, and plans were created using the new algorithm. A baseline fractionation scheme consisting of 5 fractions was compared to hypofractionated plans of 1 to 4 fractions. The effectiveness of each plan was evaluated using radiobiological criteria taken from GEC-ESTRO guidelines \cite{GECESTRO}. 

\underline{Results:} The results of this study indicate that an optimization algorithm based on biological parameters has similar functionality to traditional planning methods with the additional ability to account for fractionation effects. Using this algorithm, it was shown that plans consisting of 3 and 4 fractions have comparable target coverage with equivalent normal tissue exposure. In some specific cases, further fractionation may present acceptable target coverage as well. 

\underline{Conclusions:} An algorithm based on radiobiological parameters has clinically acceptable performance. Plans created by this optimization show that, while smaller doses per fraction result in larger target coverage with equivalent normal tissue exposure, plans consisting of 3 or 4 fractions are comparable and clinically viable. 
\end{abstract}

\maketitle

\section{Introduction}
High dose-rate (HDR) brachytherapy treatment for cervical cancer often consists of 5 or more fractions delivered over a period of several weeks \cite{cervicalfrac1, cervicalfrac2}. In contrast, recent trials for prostate cancer \cite{prostfrac1, prostfrac2, prostfrac3, prostfrac4} have been investigating the effectiveness of as few as 1 or 2 fractions of HDR brachytherapy. The difference in fractionation is due to the disparity in $\alpha/\beta$ between prostate tumors and cervical tumors, where $\alpha$ and $\beta$ are tissue-specific radiobiological parameters \cite{alphabeta}. Prostate tumor tissue and normal tissue $\alpha/\beta = 1.5-3$ Gy and are considered ``late-responding'' tissues, while most gynecological cancers have $\alpha/\beta = 10$ Gy and are considered ``early-responding'' tissues. Late responding tissues are more efficient at repairing damage and thus smaller doses per fraction preferentially spare normal tissue. Because prostate cancer has a comparable $\alpha/\beta$ to normal tissue, it does not have the same favorable reaction to lower doses per fraction and larger number of fractions per treatment. In contrast, cervical cancer is composed of early responding tissue and is not as capable at repairing damage in periods between fractionation. These effects can be seen in the classical equation for the biologically effective dose (BED), 

\begin{equation}\label{eq:BED}
    \text{BED} = D \cdot \text{RE}
\end{equation}

where the RE is the relative effectiveness of a treatment. For fractionated HDR brachytherapy, RE is given by

\begin{equation}\label{eq:RE}
    \text{RE} = 1 + \frac{d}{\alpha/\beta},
\end{equation}

where $d$ is the dose per fraction \cite{LQmodel}. Eqn. (2) indicates that, with two types of tissue, the one with a lower $\alpha/\beta$ will result in a larger BED. In the case of gynecological cancer, the tumor has a larger $\alpha/\beta$ than the surrounding normal tissue, which suggests that the same physical dose will have a larger effect in the normal tissue. This effect can be countered by reducing the dose per fraction and allowing more time between fractions for the tissue to repair itself. Tumors with a high $\alpha/\beta$ will not benefit as substantially from this treatment structure, and thus the therapeutic ratio of the treatment will increase. This indicates that dividing HDR brachytherapy treatments into a larger number of fractions will lower the toxicity to organs at risk. 

However, in some cases, hypofractionation is desired due to the increased patient comfort and lower economical impacts. Due to the non-linear nature of Eqn. (2), physical dose cannot simply be added across fractions and delivery techniques without the use of radiobiological models. Some treatment planning software allows the user to convert the final dose distribution after planning into biological dose. However, commercially available treatment planning software plans on physical dose alone and evaluation of plans are done only after the optimization is finished. In order to examine whether hypofractionation in cervical cancer can allow adequate target without increased normal tissue dose, a radiobiology-based treatment planning algorithm is needed in order to inform the optimization of fractionation effects so that plan parameters can be adjusted in order to compensate for the effects discussed above. 

Radiobiology-based treatment planning is less commonly used due to concerns such as the validity of radiobiological parameters or models and increased computation time. While there are concerns about the models that are able to be used, the algorithm built for this study can easily be updated to account for changes in these models or parameters. In addition, the computation time that is incurred by optimizing based on radiobiological parameters is not of significant proportion due to the efficient structure of the optimization considered here. Therefore, the increase in time taken to find an optimal implant is still acceptable clinically. 

While it is known that radiobiological factors affect the ability to hypofractionate HDR brachytherapy for cervical cancer, lower fractionation may be desirable clinically \cite{hypodesirable}. Using biological parameters and a radiobiology-based optimization, the effectiveness of various fractionation schemes for 6 cervical cases were examined to determine whether it is possible to achieve similar target coverage and limitation of normal tissue toxicity for various fractionation schemes.  

\section{Methods and Materials}

\subsection{Patient Demographics}
The clinical standard for cervical cancer brachytherapy at our institution is a uterine tandem and ring applicator accompanied by two interstitial needles. Image data sets from six patients previously treated at our clinic were used in this study. For each patient and implant geometry, five fractionation schemes were studied. These plans were created using an upgraded form of IPSA \cite{ipsa} that optimizes based on biological parameters.

\subsection{Optimization Algorithm}

\subsubsection{Traditional Planning}

Traditional planning at our clinic uses the IPSA inverse planning algorithm that was developed in-house and has been available commercially since 2004. Before constructing a dose plan, the patient obtains a CT scan with the catheters in place, and relevant anatomical structures and catheter positions are digitized for input into the IPSA algorithm. 

Dose points are used to control the dose distribution within the target volume and organs at risk. HDR brachytherapy produces a steep dose gradient and results in highly inhomogeneous dose distributions within the target volume and nearby organs. Thus, a large number of dose points are typically needed to control the dose within organs near active dwell positions. In order to account for this with minimal computation time, IPSA distributes dose points differently along the surface of the organs, in volumes with low doses, and volumes with high doses. Low volume dose points are uniformly distributed in the organ far away from the dwell positions. Together with the surface dose points, these points ensure appropriate target coverage and organ sparing. In addition, high volume dose calculation points are distributed to control dose hotspots around dwell positions. Available dwell positions inside the implant and within the target are assigned initial dwell times of 1 second and these times are adjusted throughout the optimization. 

After dose points and initial dwell times are assigned, a dose rate matrix is calculated based on the distance of each dose calculation point from every possible dwell position. The dose rate is calculated based on the equation outlined in TG43 \cite{tg43}: 

\begin{equation}\label{eq:doserate}
    \dot D_{ij} = S_k \Lambda \Phi_{an}(\theta, r_{ij})g(r_{ij})/r_{ij}^2,
\end{equation}

where $S_k$ is the air kerma strength, $\Lambda$ is the dose rate constant, $\Phi_{an}(\theta, r_{ij})$ is the anisotropy function, and $g(r_{ij})$ is the radial dose function. The dose rate is calculated for each dose point in all organs, and a dose rate matrix is created. The dose distribution is found from multiplying the dwell time $t$ by the dose rate $d$ due to that source at each point. To further eliminate computation time, when an adjustment is made to the dwell time of a source, the difference in time is multiplied by the dose rate and this new dose is added to the existing dose at each point,

\begin{equation}\label{eq:dose}
    D = D_0 + (t - t_0)d
\end{equation}

where $t_0$ is the previous dwell time, and $D_0$ is the previous dose. Thus, the entire dose distribution does not need to be calculated for each iteration, only the difference is found and this is added to the existing matrix.

Suitable dose plans are constructed using predefined dosimetric criteria input by the physician. A cost function is created in order to assign numerical values to the clinical criteria. Clinical objectives are defined by a range of doses that are acceptable for each organ and corresponding penalties for dose delivered outside the range. The parameters $D_L$ and $D_H$ define the low and high doses that are acceptable, and $w_L$ and $w_H$ are the corresponding weights. For the target volume, the prescription dose per fraction is usually used as the lower dose constraint on the target volume. For the organs at risk, the upper dose limit is commonly determined by a percentage of the penalty, typically half of the prescription dose. The conversion from dose to penalty, $p$, is given by 

\begin{equation}\label{eq:cost}
    p_{i} = \left\{
        \begin{array}{ll}
            -w_L(D_i - L) & \quad D_i \leq D_L \\
            0 & \quad D_L < D_i \leq D_H \\
            w_R(D_i - R) & \quad D_i > D_H,
        \end{array}
    \right.
\end{equation}

where $p_i$ is the penalty for point $i$ based on the dose, $D_i$, at that point. A total penalty is calculated for each organ by summing the penalties over each point in an organ: 

\begin{equation}\label{eq:penalty}
    P = \sum_{i}^{N} = \frac{p_i}{N}.
\end{equation}

The cost function for the dwell time configuration of a given iteration is the sum of the penalties for each organ. The algorithm then systematically alters dwell times for each position in order to minimize the total penalty using a simulated annealing optimization engine where ideal cooling parameters have previously been found \cite{ipsa}. 

\begin{figure}\label{fig:workflow}
    \begin{center}
        \begin{adjustbox}{width=0.48\textwidth,height=0.48\textheight,keepaspectratio}
            \begin{tikzpicture}[node distance = 2cm, auto]

                \node (in1) [input] {\small{Radiobiological dosimetric criteria and digitized structures}};
                \node (pro0) [process, below of=in1] {\small{Dose calculation points created}};
                \node (pro1) [process, below of=pro0] {\small{Initial dwell times assigned}};
                \node (pro2) [process, below of=pro1] {\small{Dose matrix calculated}};
                \node (pro3) [process, below of=pro2] {\small{Dose distribution calculated}};
                \node (pro4) [process, below of=pro3] {\small{Biological dose matrix calculated}};
                \node (dec1) [decision, below of=pro2, yshift=-4.8cm] {\small{EQD2 Penalty evaluation}};
                \node (out1) [input, below of=dec1, yshift=-0.8cm] {Optimal dwell time configuration};
                \node (pro2a) [process, right of=pro4, xshift=2cm] {Dose distribution adjustments};
                \node (pro2b) [process, right of=dec1, xshift=2cm] {Dwell time adjustments};

                \draw [arrow] (in1) -- (pro0);
                \draw [arrow] (pro0) -- (pro1);
                \draw [arrow] (pro1) -- (pro2);
                \draw [arrow] (pro2) -- (pro3);
                \draw [arrow] (pro3) -- (pro4);
                \draw [arrow] (pro4) -- (dec1);
                \draw [arrow] (dec1) -- (out1);
                \draw [arrow] (dec1) -- (pro2b);
                \draw [arrow] (pro2a) |- (pro3);
                \draw [arrow] (pro2b) -- (pro2a);
    
            \end{tikzpicture}
        \end{adjustbox}
    \end{center}
    \caption{Algorithmic steps followed by the radiobiological optimization used in this study.}
\end{figure}
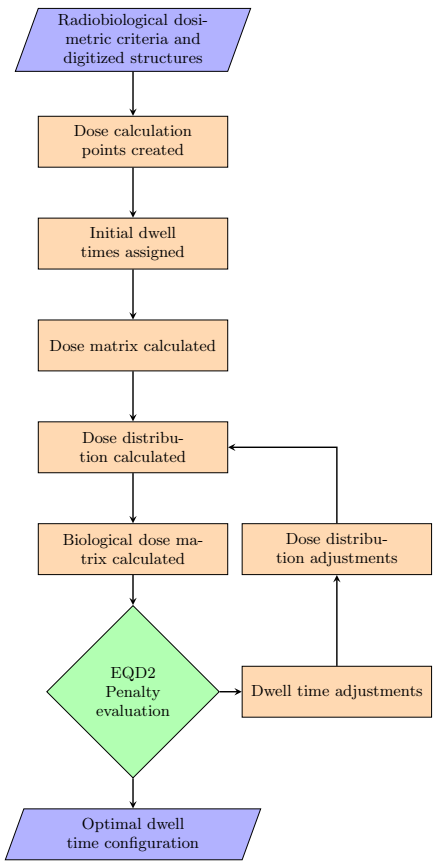

\subsubsection{Biological Planning}

The radiobiological optimization used in this study alters and adds several steps in the traditional optimization. A workflow diagram of the new optimization can be seen in Fig. 1. The patient data acquisition is consistent between the two algorithms, but the dose criteria used differs. In traditional panning, the dose limits are based on physical prescription dose for one fraction. The optimization built for this study uses criteria based on the entire treatment and uses biological models to determine the effectiveness of the complete plan using all fractions. For this study the radiobiological dose used was equivalent dose in 2 Gy fractions (EQD2) which is converted from BED using

\begin{equation}\label{eq:EQD2}
    \text{EQD2} = \frac{\text{BED}}{1 + \frac{2}{\alpha/\beta}}, 
\end{equation}

where BED is found using Eqn. (1). The acceptable doses $D_L$ and $D_H$ are taken from GEC-ESTRO recommendations for the total amount of dose that should be delivered across all fractions \cite{GECESTRO}. Because the doses are given in EQD2, there is no need to change the dosimetric criteria based on the number of fractions in a treatment. Traditionally, the number of fractions would be decided by a physician before optimization. The prescription dose for each fraction would then depend on the total number of fractions so that, regardless of the fractionation scheme, the total biological dose delivered at the end of treatment is about the same. Because the optimization developed in this study accounts for all fractions, individual fraction prescriptions do not need to be determined, and the optimization finds the ideal dose per fraction so that organs at risk are spared. 

The parameters $D_L$ and $D_H$ remain constant for each patient and fractionation scheme, and the user inputs the number of fractions. The weights $w_L$ and $w_H$ can be adjusted by the physician in the same manner as before if the dose distribution is not satisfactory after optimization. 

After all dosimetric criteria has been input, the optimization assigns dwell times and dose points in the same manner as the traditional IPSA algorithm. The dose rate matrix is based on physical dose and is calculated using Eqn. (3). The initial physical dose distribution is then used to create an additional biological dose distribution matrix. The total dose at each point from all dwell positions is converted into EQD2 using Eqn. (6) and stored in the new matrix. When the physical dose is converted to EQD2, the number of fractions is taken into account, where the conversion assumes that the implant and dwell time configuration will be the same for each fraction. This matrix is then the one used to calculate the cost function rather than the original physical dose distribution. Penalties are calculated in the same manner as Eqn. (4) and summed over each organ. 

Based on the penalties accumulated, the algorithm adjusts the dwell time configuration, and differences in dwell times are calculated for each dwell position. Due to the non-linear nature of the linear quadratic model, the dose difference due to the change in dwell time cannot be simply added to the previous dose distribution as is done in the physical optimization. The dose change is added to the physical dose distribution and the biological dose distribution is then re-calculated from the new physical dose distribution matrix. The process then repeats until a maximum is located or the maximum number of iterations is reached.

In order to determine the validity of this optimization, the output of both optimizations were compared. For the biological optimization to be properly functional, a dose-volume histogram made from the resulting dose distribution should mirror the output of the traditional optimization in shape and scale. Dose-volume histograms were created using the physical dose distribution from both optimizations and it was seen that the two histograms had similar shape and structure. Dose-volume histograms from the original IPSA and the radiobiology-based IPSA optimization can be seen in Fig. 2 which indicates that the two optimizations produce similar, but varying dose distributions. A small deviation in dose distribution is evidence that the optimization adjusts for fractionation effects; however, a similar dose distribution to those typically seen clinically should be kept in order to obtain the desired clinical results. In addition, the dose distribution output by the physical optimization was converted into EQD2, and this produced similar results. This indicates that the distribution created by the biological optimization is reasonable but different, showing that informing that optimization about biological parameters promotes a change in configuration. 

\begin{table}\label{table}
\setlength{\tabcolsep}{1.75em}
\centering
\vspace{0.25in}
\begin{tabular}{lll}
\hline
\multicolumn{3}{c}{Biological Parameters} \\
\hline
Organ & Guideline Dose & EBRT Dose \\
\hline
Rectum & 65 Gy &    45 Gy\\
Bladder & 70 Gy &    45 Gy\\
Bowel & 65 Gy &    45 Gy\\
CTV & 85 Gy &    55 Gy \\
\hline
\end{tabular}
\caption{Biological parameters used to compare different fractionated treatments. Recommended values were taken from GEC-ESTRO guidelines \cite{GECESTRO}.}
\end{table}

\begin{figure}\label{fig:dvh}
\includegraphics[scale=0.55]{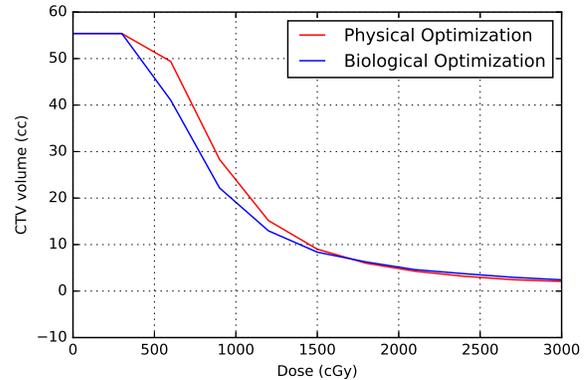}
\caption{Dose-volume histogram created from physical dose distribution using the original IPSA and radiobiology-based IPSA optimization for case A in this study.}
\end{figure}

\subsection{Plan Evaluation}
The quality of each plan was judged based on the biological dose delivered to 2cc of each organ at risk (D2cc). While creating the plans, the weights on the organs at risk and the target volume were adjusted so that the dose to the organs just approached the maximum recommended values. The maximum values for the organs at risk were taken from GEC-ESTRO recommendations for cervix cancer brachytherapy which determined that the organs at risk should receive less than 65-70 Gy (EQD2) after the completion of all fractions and that 90$\%$ of the target volume should receive 85 Gy (EQD2). Total EQD2 recommendations are given as total doses including both external beam and brachytherapy and have not been thoroughly investigated for either as monotherapy. The focus of this study was on the dose distribution due to brachytherapy alone, and thus a typical external beam dose was assigned to each of the organs at risk and target. The typical doses from external beam for the organs at risk and the target can be found in Table 1 and were taken from GEC-ESTRO guidelines \cite{GECESTRO}. This was done in order to ensure that the toxicity on the organs at risk all remained under the recommended values and allowed for consistency between the plans. For the organs at risk that were not near the maximum recommended values, a constant toxicity was kept between fractionation schemes. 

\begin{table*}\label{table}
\setlength{\tabcolsep}{1.75em}
\centering
\vspace{0.25in}
\begin{tabular}{llllll}
\hline
\multicolumn{6}{c}{Physical Dose to Target} \\
\hline
Case & 1 fraction & 2 fractions & 3 fractions & 4 fractions & 5 fractions\\
\hline
A & 15 Gy & 12 Gy & 9 Gy & 9 Gy & 6 Gy\\
B & 15 Gy & 12 Gy & 9 Gy & 9 Gy & 6 Gy\\
C & 18 Gy & 12 Gy & 9 Gy & 9 Gy & 6 Gy\\
D & 18 Gy & 15 Gy & 12 Gy & 9 Gy & 9 Gy\\
E & 12 Gy & 9 Gy & 9 Gy & 6 Gy & 6 Gy\\
F & 15 Gy & 12 Gy & 12 Gy & 9 Gy & 6 Gy\\
\hline
\end{tabular}
\caption{Physical dose delivered to 90$\%$ of the target volume.}
\end{table*}

In addition to calculating the EQD2 values, the equivalent uniform biologically effective dose, or EUBED, was compared for each case. EUBED takes into account the inhomogeneity in dose distribution and is important for this study due to the effect that high doses per fractions can have on organs \cite{EUBED, EQD2good}. Hot spots in tissue can lead to higher toxicity and are more prevalent in plans with fewer fractions. The EUBED for an entire organ is calculated by considering the BED of each voxel and combining each voxel to give an effective dose for the entire organ. Thus, areas of high dose are taken into consideration when using EUBED and can more accurately represent the toxicity to organs at risk. The EUBED is found using the equation from Jones and Hoban [12],

\begin{equation}\label{eq:EUBED}
    \text{EUBED} = -\frac{1}{\alpha} \text{ln} \left(\sum_{i}^{N}v_i e^{\alpha \text{BED}_i}\right)
\end{equation}

where $\text{BED}_i$ is the BED from each individual voxel. The weights on each organ were adjusted to give equal D2cc, and the EUBED was recorded and compared across fractionation schemes.

The percent volume of the CTV that received 85 Gy (EQD2) was recorded for each case and fractionation scheme. Loss of dose to the target area is commonly a result of organ at risk sparing in order to reduce toxicity. Thus, we examine the amount of dose that is able to be delivered to the target while exposing the organs at risk to the same amount of biologically effective dose across all fractions.

\section{Results}

The volume of the target that received the recommended 85 Gy for each case across the 5 fractionation schemes is shown in Fig. 3. In each case, as the number of fractions increases, the target coverage also increases. The plans with a greater number of fractions consistently show more target coverage with the same amount of dose to the organs at risk due to the dose-sparing effect that was discussed in the introduction. However, for cases C and D, the volume of target coverage is fairly flat across all fractionation schemes, indicating that increasing the number of fractions does not allow for substantially more target coverage. For cases A, B, and E, the trend shows a significant increase in target coverage corresponding to an increase from 1 to 3 fractions. Thus, there is a considerable benefit to increasing the number of fractions in this range. In the range above 3 fractions, the increase in target coverage substantially decreases and approaches a flat trend. Thus, in these cases, the number of fractions could be reduced from 5 to 3 without considerable loss in coverage. Therefore, while 1 or 2 fractions may not be feasible in most cases for cervical HDR brachytherapy, fractionation schemes consisting of 3 and 4 fractions are clinically viable. In addition, for particularly well placed implants, it may be possible to reduce the number of fractions even further without significant loss of target coverage.

\begin{figure}\label{fig:main}
\centering
\includegraphics[scale=0.6]{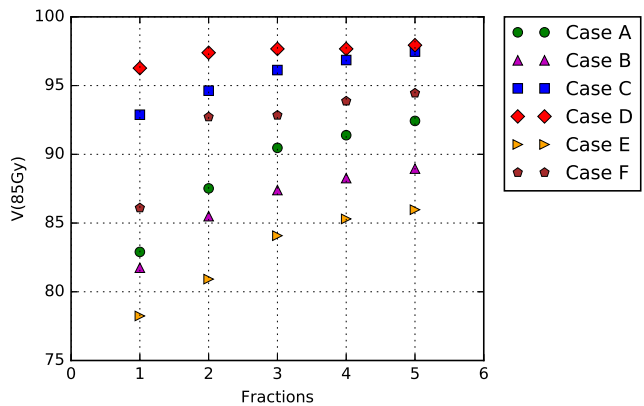}
\caption{Percent of CTV receiving 85 Gy versus the number of fractions in the total treatment for each case discussed in this study.}
\end{figure}

The EUBED for the bladder for each fraction is shown in Fig. 4. The plans consisting of fewer fractions show a higher EUBED that results from the higher degree of inhomogeneity that results from hypofractionation discussed earlier. Although the EUBED increases with fewer fractions, the increase is not of significant proportion. This indicates that lower fraction plans are not suffering from a much larger increase in hot spots within the organs and that toxicity between plans should be similar. 

\begin{figure}\label{fig:eubed}
\centering
\includegraphics[scale=0.6]{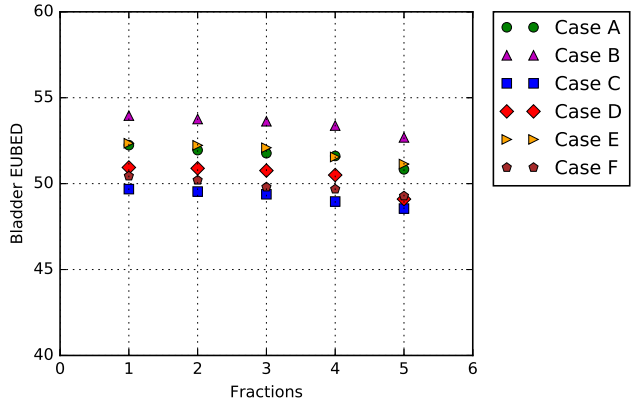}
\caption{EUBED of the bladder for each fraction. Similar structure can be seen for the other organs at risk considered.}
\end{figure}

Because each of these plans were created with the upgraded version of IPSA that optimizes based on BED alone, prescription doses per fraction did not need to be assigned. For reference, the equivalent prescription dose that would have been assigned based on the dose distribution is shown in Table 2. These values were found by calculating the physical dose that was delivered to 90$\%$ of the target volume after the dwell times and positions were optimized based on BED.

\section{Discussion}
There is a desire for hypofractionated treatments for cervical cancer due to patient experience and economical impact \cite{hypodesirable}. If HDR brachytherapy could be reduced to only one fraction, the treatment could become an outpatient procedure, significantly reducing patient discomfort and time. While this research has shown single fraction HDR brachytherapy treatments are not viable in most cases, plans consisting of fewer fractions are clinically acceptable and can be used to reduce the number of implants needed. 

Using the new algorithm that was created for this study, physicians can plan in real time and use the information given from the optimization to decide the number of fractions needed for a specific patient. From the results of this study, it can be seen that the ability to deliver fewer fractions depends on both the quality of the implant, and the patient-specific anatomy. In a particularly well-placed implant, such as in cases C and D, the physician could decide to schedule fewer than 5 fractions from the results of the optimization. In the same manner, for cases such as A, B, and E, the physician could determine that 3 or more fractions are needed and schedule further implants.  By incorporating the biology-based treatment planning algorithm, physicians can determine the quality of a specific implant and use this information to determine the appropriate number of fractions for a given patient.

\def\bibsection{\section*{\refname}} 

\bibliography{references}
\bibliographystyle{ieeetr}

\end{document}